\documentclass[onecolumn]{IEEEtran}

\usepackage{amsfonts,amsmath,amssymb,amsthm}
\usepackage{balance}
\usepackage{graphicx}
\usepackage{amsmath}
\usepackage{graphicx}
\usepackage{calc}
\usepackage{epstopdf}
\usepackage{subfigure}
\usepackage[T1]{fontenc}
\usepackage{amsfonts}
\usepackage{bm}
\usepackage{bbm}
 \usepackage[ruled,norelsize]{algorithm2e}
\usepackage{algorithmic}
\usepackage{verbatim}
\usepackage{amssymb}
\usepackage{amsthm}

\usepackage{color}
\usepackage{subfigure}
\usepackage{array}
\usepackage{cite}
\usepackage{stfloats}
\begin{document}

\title{A Joint Resource Allocation Scheme for Multi-User Full-Duplex OFDMA Systems}
\author{
	\IEEEauthorblockN{Yang You\IEEEauthorrefmark{1},  Chong Qin \IEEEauthorrefmark{2} and Yi Gong\IEEEauthorrefmark{2}}

\IEEEauthorblockA{\IEEEauthorrefmark{1}School of Electrical Engineering, KTH Royal Institute of Technology, Stockholm, Sweden} \\

\IEEEauthorblockA{\IEEEauthorrefmark{2}Department of  Electronic Engineering, South University of Science and Technology, Shen Zhen, China} 

}
\maketitle
\begin{abstract}
Full-duplex is a promising technology and great system performance enhancement can be achieved especially when we apply it to base stations. In this paper, we consider a Multi-User OFDMA network consisting of one Full-duplex base station, several uplink users, and a number of downlink users. A joint subcarrier, downlink user and uplink user mapping and power allocation optimization problem is investigated. In detail, we first decompose the 3-dimensional mapping problem into three 2-dimensional sub-problems and solve them by iteratively using classical Hungarian Method. Then, based on the dual method, we sequentially solve the power allocation and 3-dimensional mapping problem at each dual point. And the optimal solution to the dual problem is derived by using sub-gradient method. Unlike existing methods that only solve one sub-problem but with a high computation complexity, we tackle both of them and successfully reducing computation complexity from exponential to polynomial order. Numerical simulations are conducted to verify the proposed system.
\end{abstract}

 \begin{IEEEkeywords}
 	Full-duplex Base Station (FD-BS), resource allocation, 3-dimensional binary assignment, Lagrange dual method
 	 \end{IEEEkeywords} 
\section{Introduction}
Full duplex (FD) is a promising technology as it is able to highly improve the spectrum efficiency by transmitting and receiving signals within the same frequency band. Due to the effect of Self-interference (SI), most of the previous research were restricted to Half-duplex (HD). Recently, as mentioned in [1], FD communication came to the stage since great progress had been made in developing SI cancellation schemes. In particular, FD relaying transmission technology is widely considered in the cooperative communication scenarios, such as [2-4]. However, FD technology is not only applicable for the relay node, it can also be applied to the Base Station(BS) as well. Currently, there are few research work corresponding to FD-BS. For instance, [5] investigates the extra benefits on degrees of freedom of cellular networks brought by the FD BS and [6] considers the security problem of a system with a more advanced BS.

Specifically, the FD-BS enables the system to communicate with uplink and downlink users in the same band simultaneously and an evident increase in system throughput can potentially be achieved. In order to optimize the system performance, a low-complexity and reasonable resource allocation strategy needs to be investigated in the FD-BS scenario. There exists some work related with this but failing achieving the goal, e.g. [7] and [8].  In [7], there is a cellular network consisting of one FD-BS, several users and sub-channels. A joint user scheduling, power control and channel assignment scheme is taken into account, but no closed-form solution is provided. However, a potential solution for power allocation problem is given in [8], but the authors are only considering the mapping either between downlink user and subchannel or between uplink user and subchannel.

In this letter, we revisit the scenario mentioned in [3] and [8], majoring in investigating a low complexity 3-dimensional pairing problem between Uplink User (UUE), Downlink User (DUE) and Subchannel, which is a combinatorial optimization problem. As well as the power allocation problem in a OFDMA cellular system with FD-BS. The contributions of our proposed approach are twofold: (1) providing solving method for the power allocation problem for FD-BS system (2) significant complexity reduction on 3D pairing problem by proposed assignment scheme. Specifically, we divide the problems into two sub ones. Given a possible mapping between users and subchannel, the well-known Lagarange dual method is provided for resolving the power allocation problems. Then decompose the 3D pairing problem into three 2D paring subproblems and solving it iteratively it using classical Hungarian algorithm. As shown in [9], the result will at least converge to a local optimal solution after using Hungarian method at most 5 times, which leads to a computation complexity in polynomial order. Besides, numerical simulations  strongly approve the correctness of our proposed method.

The rest of the paper is organized as follows. Section II presents the system model. Sections III proposes a low complexity multi-dimensional Hungarian method for solving the binary assignment problem, and Section IV provides the joint power allocation and binary assignment solution. The numerical results are presented in Sections V. Finally Section VI draws some conclusions.

\section{System Model and Problem Formulation}

\subsection{System Model}
As shown in Fig.1, we consider a single cell multi-user FD-BS OFDMA system and it consists of one FD mode base-station (BS), M HD mode uplink users
and N HD mode downlink users. In particular, they are all equipped with single antenna. The total bandwidth is divided into $K$ rayleigh flat fading subchannels, and each of them can only be shared by only one uplink and downlink user pair $(m,n)$, forming the 3-dimensional combination $(m,n,k)$. Due to the imperfect SI cancellation, the BS will suffer the SI from itself when it receives signal from the uplink user, and the SI power at BS receiver is generally modified as an additive white Gaussian noise (AWGN).

\begin{figure}
  \centering
  \includegraphics[width=9cm]{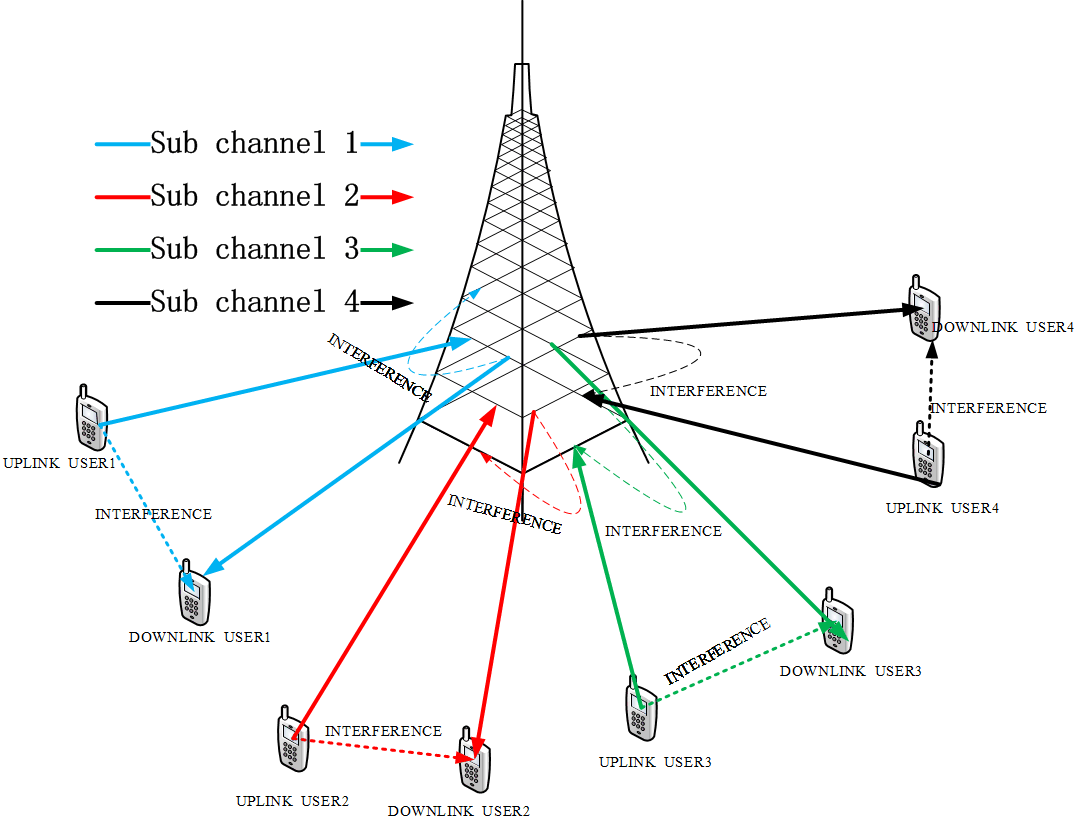}\\
  \caption{FD-BS system}\label{PCF}
\end{figure}

\subsection{Problem Formulation}
Define $m\in\{1,....M\}$, $n\in\{1,....N\}$ and $k\in\{1,....K\}$ as the index of UUE, DUE, and Subchannel. Consequently, using $P_{m,b}^k$ and $P_{b,n}^k$ to denote UUE $m$ transmit power for BS over subcarrier k and BS transmit power for DUE $n$ over subcarrier k repectively, while the notation $h_{m,b}^k$ and $h_{b,n}^k$ represent the corresponding channel coefficient. In this case, the received signal at BS receiver and DUE $n$ can be written as:

\begin{equation} \label{eq:Eq1}
y_{m,b}^k=\sqrt{P_{m,b}^k} h_{m,b}^k x_m+Z_D+Z_B
\end{equation}
\begin{equation} \label{eq:Eq1}
y_{b,n}^k=\sqrt{P_{b,n}^k} h_{b,n}^k x_n+\sqrt{P_{m,b}^k} h_{m,n}^k x_m+Z_B
\end{equation}
where $ \bf{\sl{E}}\{\left| x_m \right|^2\}=\bf{\sl{E}}\{\left|x_m\right|^2\}=1$, and $ Z_D \sim CN(0,\sigma_D^2) $ and $ Z_B \sim CN(0,\sigma_B^2) $ denote the self-interference and receiver noise, respectively. By Shannon formula, the corresponding uplink and downlink throughput over the subchannel $k$ can be  expressed as:

\begin{equation} \label{eq:Eq1}
R_{m,b,k}^{up}=\log \left(1+\frac{P_{m,b}^k \left| h_{m,b}^k \right|^2}{\sigma_D^2+\sigma_B^2} \right)
\end{equation}

\begin{equation} \label{eq:Eq1}
R_{b,n,k}^{down}=\log \left(1+\frac{P_{b,n}^k \left| h_{b,n}^k \right|^2}{P_{m,b}^k \left| h_{m,n}^k \right|^2+\sigma_N^2} \right)%
\end{equation}

Hence, the sum rate of transmission pair $(m,n,k)$ is:
\begin{equation} \label{eq:Eq1}
R_{m,n}^k = R_{m,b,k}^{up} + R_{b,n,k}^{down}
\end{equation}

Provided (3), (4) and (5), to maximize the system throughput, the joint 3D mapping and power allocation problem is now formulated as:
\begin{equation}
\begin{aligned}
&(P1):\mathop {\max }\limits_{p_{b,n}^k, p_{m,b}^k, X} \sum\limits_{m = 1}^M {\sum\limits_{n = 1}^N {\sum\limits_{k = 1}^K {x_{m,n}^kR_{m,n}^k} } }\\
&\qquad  \qquad ~~ s. t. ~~~c1:\sum_{n=1}^{N}\sum_{k=1}^K p_{b,n}^k \leq P_b \\
&\qquad  \qquad ~~~~~~~~~ c2:\sum_{k=1}^K P_{m,b}^k \leq P_m, \forall m \\
&\qquad  \qquad ~~~~~~~~~  c3: X_{m,n}^k=\{0,1\} \\
&\qquad  \qquad ~~~~~~~~~  c4:\sum\limits_{m = 1}^M \sum\limits_{n = 1}^N x_{m,n}^k = 1,\forall k;\\
&\qquad  \qquad ~~~~~~~~~~~~~  \sum\limits_{m = 1}^M\sum\limits_{k = 1}^K {x_{m,n}^k} = 1,\forall n; \\
&\qquad  \qquad ~~~~~~~~~~~~~  \sum\limits_{n = 1}^N\sum\limits_{k = 1}^K {x_{m,n}^k} = 1, \forall m;
\end{aligned}
\end{equation}

Where $X$ is the $M\times N \times K$ 3-dimensional assignment matrix with $x_{m,n}^k=1$ if subchannel $k$ is assigned to UUE-DUE pair $(m,n)$, and $x_{m,n}^k=0$, otherwise. The constraint c4 follows the fact each subchannel can only be assigned to one UUE-DUE pair, and constraint c1, c2 indicates the individual power constraint for UUE and BS, respectively.

\section{low-complexity 3-dimensional binary assignment scheme}
In this section, we assume a fixed power allocation scheme, and (P1) can be simplified to the following with a fixed throughput ${R_{m,n}^k}^*$ for each $(m,n,k)$ pair,
\begin{equation} \label{eq:Eq1}
\begin{aligned}
&\mathop{\max }\limits_{X}\sum\limits_{m = 1}^M \sum\limits_{n = 1}^N {\sum\limits_{k = 1}^K {x_{m,n}^k{R_{m,n}^k}^*} }\\
&~~ s.t.~~c3,c4
\end{aligned}
\end{equation}

which is a NP-complete problem, with the optimal solution to be exhaustive searching. According to \cite{ex}, the complexity of exhaustive search is in exponential order for the proposed optimization problem, which can not be solved in polynomial time. However, since it's proved in \cite{D2D}, the 3-dimensional mapping problem can be solved by using 2-dimensional Hungarian method in 5 iterations with near-optimal performance. And the complexity turns to be $O(5L^{3})$ where $L=\max( M,N,K)  $, which is polynomial and much lower than the optimal scheme.

In this case, we can decompose our 3D mapping problem into following three different 2D mapping subproblems and solve them iteratively,

1). 2D mapping between the UUE-DUE pair $(m,n)$ and Subchannel $k$, the solution is $X_{1}^*$.

2). 2D mapping between the UUE-Subchannel pair $(m,k)$ and DUE $n$, the solution is $X_{2}^*$.

3). 2D mapping between the DUE-subchannel pair $(n,k)$ and UUE $m$, the solution is $X_{3}^*$.

To initialize the scheme, set an arbitrary mapping matrix $X_{M,N,K}$ which satisfies constraint c3 and c4. Then, define a 2-dimensional index including indices of UUE $m$ and DUE $n$ be $\chi_{0}= \{(m,n)\mid x_{m,n}^k =1,\forall m,n,k\}$. Here, for simplicity, we use $d$ to represent the pair $(m,n)$, when $M\leq N$, $d=m$; and $d=n$ otherwise.

With the pair $(m,n)$, the two different dimensions UUE $m$ and DUE $n$ could be regarded as a joint dimension with index $d$, then we could adopt a 2-dimensional mapping matrix $T_{\min(M,N)\times K}= [{t_{d}}^k]$ with ${t_{d}}^k=1$ if subchannel $k$ is allocated to UUE-DUE pair $(m,n)$, otherwise ${t_{d}}^k=0$.

In this case, the mapping relationship shown in the initial mapping matrix $X_{M,N,K}$ could be demonstrated by the set $(\chi_{0},T_{0})$. And the problem turns to be searching an optimal subchannel assignment matrix ${T_{0}}^*$ for the current UUE-DUE pair set $\chi_{0}$, which can be formulated by the following,

\begin{equation} \label{eq:Eq1}
\begin{aligned}
&\mathop {\max }\limits_{{T_{0}}} \sum\limits_{d = 1}^{\min(M,N)} {\sum\limits_{k = 1}^K {T_{d}^k{R_{d,k}^*}} }\\
&~~~s.t. \sum\limits_{d = 1}^{\min(M,N)}{t_{d}^k = 1},\forall k;\\
&~\quad \qquad R_{d,k}^*= {R_{m,n}^k}^*, \forall (m,n)\in \chi_{0}
\end{aligned}
\end{equation}

The classic Hungarian method can be used to solve the above problem, yielding a new 3D mapping matrix $X_{1}^*$ which has the same mapping relationship with set $(\chi_{0},{T_{0}}^*)$. And this $X_{1}^*$ can be set as the initial mapping matrix for subproblem 2). By applying same strategy iteratively on the three subproblems, we can derive the optimal solutions for each iteration in the sequence $X_{1}^*\rightarrow X_{2}^*\rightarrow X_{3}^*\rightarrow X_{4}^*\rightarrow X_{5}^*$. $X_{5}^*$ can be chosen as the global optimal 3D binary mapping matrix.

\section{Joint power allocation and 3D mapping}

In this section, we further consider the power allocation problem. As the joint power allocation and binary assignment
problem is a non-convex mixed combinatorial problem, which is extremely complicated to solve. We propose the following method by solving the Lagrange dual problem instead of solving the primal problem, since it has been shown in \cite{dual}, for multicarrier systems, the duality gap of a non-convex resource allocation problem is negligible when the number of subcarriers becomes sufficiently large.

To combine the integer constraint with power constraint. We define the following virtual power,

\begin{equation}
P_{m,n}^{k,up}=P_{m,b}^k x_{m,n}^k
\end{equation}
\begin{equation}
P_{m,n}^{k,down}=P_{b,n}^k x_{m,n}^k
\end{equation}

With (9) and (10), constraint c1 and c2 can be transformed to,
\begin{equation}
c5: \sum_{k=1}^K\sum_{n=1}^N P_{m,n}^{k,up}\leq P_m \qquad\forall m
\end{equation}
\begin{equation}
c6: \sum_{k=1}^K\sum_{n=1}^N\sum_{m=1}^MP_{m,n}^{k,down}\leq P_b
\end{equation}

Consequently, (P1) can be transformed to the following optimization problem (P5),

\begin{equation} \label{eq:Eq1}
\begin{aligned}
&(P3): \mathop{\max }\limits_{{X,\overline{P}}}\sum\limits_{m = 1}^M \sum\limits_{n = 1}^N {\sum\limits_{k = 1}^K {\overline{R_{m,n}^k}} }\\
&~~~~~~~~~s.t.~~c3,~c4,~c5~\text{and}~c6
\end{aligned}
\end{equation}

Where $\overline{P}=\left( P_{m,n}^{k,up},P_{m,n}^{k,down} \right)$, and the throughput defined by,
\begin{equation}
\overline{R_{m,n}^{k,up}}=\log \left(1+\frac{P_{m,n}^{k,up} \left| h_{m,b}^k \right|^2}{\sigma_D^2+\sigma_B^2} \right) \\
\end{equation}
\begin{equation}
\overline{R_{b,n}^{k,down}}=\log \left( 1+\frac{P_{m,n}^{k,down} \left| h_{b,n}^k \right|^2}{P_{m,n}^{k,up} \left| h_{m,n}^k \right|^2+\sigma_N^2} \right)
\end{equation}
\begin{equation}
\overline{R_{m,n}^k}=\overline{R_{m,n}^{k,up}}+\overline{R_{b,n}^{k,down}}
\end{equation}

In this case the dual function can be written as,

\begin{equation}
g(\lambda)=\mathop{\max}\limits_{X,\overline{P}}L(\overline{P},\lambda)
\end{equation}

where the Lagarange function and dual vector $\lambda$ is given by ,
\begin{equation}
\begin{aligned}
&L(\overline{P},\lambda)=\sum\limits_{m = 1}^M \sum\limits_{n = 1}^N \sum\limits_{k = 1}^K (\overline{R_{m,n}^k}-\lambda_m P_{m,n}^{k,up}-\lambda_b P_{m,n}^{k,down})\\
&~~~~~~~~~~~~~ +\sum_{m=1}^M\lambda_m P_m
  +\lambda_bP_b\\
&\lambda=\{\lambda_1,...,\lambda_M,\lambda_b\}
\end{aligned}
\end{equation}

And (p2) can be solved by solving its dual optimization problem,

\begin{equation}
\begin{aligned}
&\mathop{\min}\limits_{\lambda}~~g(\lambda)\\
&~~s.t.~~ \lambda \geq 0
\end{aligned}
\end{equation}

The proposed dual optimization problem (19) can be solved by firstly solving (17) at each given dual point and updating the dual function using its subgradient. The subgradient of $g(\lambda)$ can be obtained by using a similar method as mentioned in \cite{dual}. And the dual variables can be updated based on the expression (20)-(21), where $[\bullet]^{+}$ denotes $\max(0,\bullet)$ and the step size $\pi^{(l)}$ follows the diminishing policy in \cite{step}, i.e., $\pi^{(l)}=\pi^{(0)}/\sqrt{l}$.

\begin{equation}
\lambda_b^{(l+1)}=\left[\lambda_b^{(l)}-\pi^{(l)}\left( P_b-\sum_{m=1}^M\sum_{n=1}^N\sum_{k=1}^KP_{m,n}^{k,down} \right)\right]^{+} \\\\
\end{equation}

\begin{equation}
\lambda_m^{(l+1)}=\left[\lambda_m^{(l)}-\pi^{(l)}\left( P_m\\-\sum_{n=1}^N\sum_{k=1}^KP_{m,n}^{k,up}\right)\right]^{+} \quad\forall m \\\\
\end{equation}

To compute the dual function at each given dual point, we need to find the optimal mapping matrix $X$ and the optimal power allocation vector $\overline{P}$. In detail, an optimal power allocation scheme is derived given each possible mapping pair $(m,n,k)$, with the solution to power allocation we can do 3D mapping using the scheme proposed in Section III.

Fixing the pair $(m,n,k)$, and defining $a_{m,b}^k=\frac{\left| h_{m,b}^k \right|^2}{\sigma_D^2+\sigma_B^2}$, $a_{b,n}^k=\frac{\left| h_{b,n}^k \right|^2}{\sigma_N^2}$ and $a_{m,n}^k=\frac{ \left| h_{m,n}^k \right|^2}{\sigma_N^2}$ the power allocation problem can be formulated into,

\begin{equation}
\begin{aligned}
&P3:\mathop {\max }\limits_{\left\{ \overline{P} \right\}}~~
{\log(1+P_{m,n}^{k,up}a_{m,b}^k)+\log(1+\frac{P_{m,n}^{k,down}a_{b,n}^k}{P_{m,n}^{k,up}a_{m,n}^k+1})}\\
&\qquad \qquad \ \quad -\lambda_m P_{m,n}^{k,up}-\lambda_bP_{m,n}^{k,down}\\
&\qquad~~  \ s.t.~~ \overline{P}=\left[P_{m,n}^{k,up},P_{m,n}^{k,down}\right]\geq 0
\end{aligned}
\end{equation}

With (22), by first applying Karush-Kuhn-Tucker (KKT) conditions then considering the different feasible regions, we can derive the optimal power allocation solution as shown in Appendix A.

\section{Numerical Results}
In this section, we present the simulation results to demonstrate the performance of the proposed algorithm. We consider
a cell with radius 200 m, 8 UUE and 8 DUE are generated randomly within the same cell. To simulate practical channel
propagation, the LTE typical urban channel model is employed. The spectral density of noise is $-126$ dBm/Hz and the total bandwidth is $180$ KHz shared by $64$ subchannels. The self-interference is modeled as AWGN, with the power 3 dB larger than the noise power. And the peak power constraints for all the uplink users are the same and set to be 5 dB lower than the maximum BS transmit power.

\begin{figure}[h]
 \centering
  \includegraphics[width=9cm]{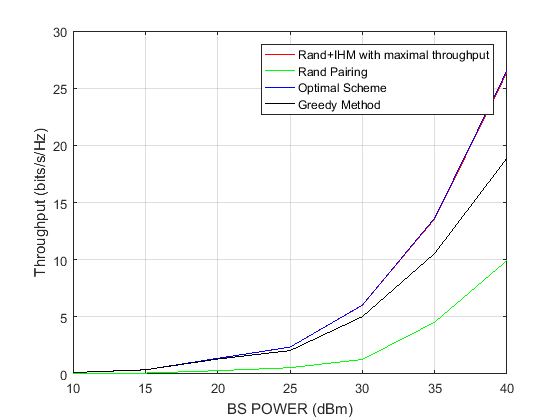}\\
  \caption{Performance comparison between different binary assignment schemes}\label{PCF}
\end{figure}

\begin{figure}[h]
 \centering
  \includegraphics[width=9cm]{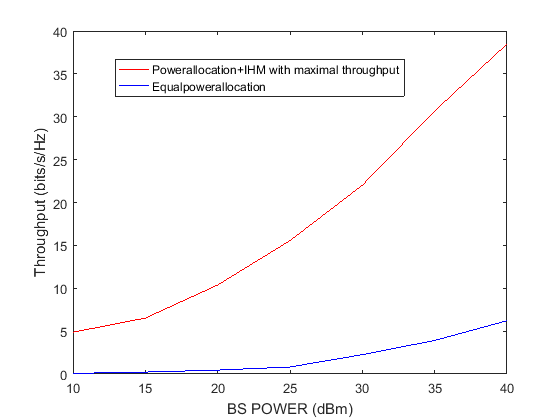}\\
  \caption{Performance of proposed joint resource allocation scheme compared with equal power allocation}\label{PCF}
\end{figure}

\begin{figure}[h]
 \centering
  \includegraphics[width=9cm]{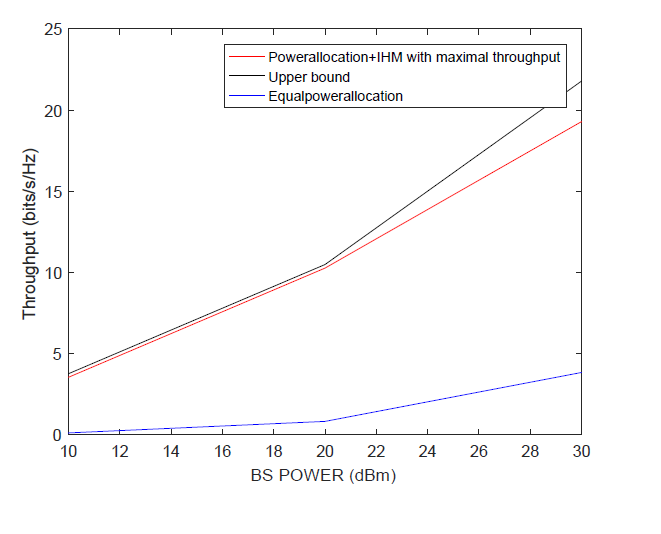}\\
  \caption{Upper bound of the average sum-rate}\label{PCF}
\end{figure}

\subsection{Binary Assignment Scheme Comparison}

With the maximum transmit power of UUE to be equal power allocation scheme, compare our proposed 3D mapping scheme with the following three benchmark schemes,

1). Exhaustive searching scheme, which is optimal.

2). Random mapping scheme, which generates a random 3D mapping solution in each iteration.

3). Greedy algorithm, each UUE $m$ select the DUE-subchannel pair $(n,k)$, which can maximize the throughput of pair $(m,n,k)$. Once the pair $(n,r)$ is selected, the other UUE can't select it any more.

As is shown in Fig.2, with BS peak power constraint varying from 10 dBm to 30 dBm, our proposed 3D mapping scheme almost has the same performance with the optimal scheme, while the gaps between the other approaches and the optimal scheme are quite large. Besides, we can see from the figure, since we apply equal power allocation, it will lead to a growing SINR difference between different $(m,n,k)$ pairs when the total power constraint is getting larger, which means a growing gap between different mapping results.

\subsection{Joint Scheme Comparison}

By comparing our proposed joint resource allocation scheme with the 3D mapping scheme based on equal power allocation, we can see from Fig.3, our power allocation scheme significantly enhance the system performance as it allocate more power to the $(m,n,k)$ pairs with good link quality.

To show the close-to-optimal performance of our proposed scheme, we introduce a power allocation scheme which exhaustively search the optimal solution. Due to the huge complexity of exhaustive search, we reduce the total number of users to $8$, which means 4 uplink users and 4 downlink users. The performance of our proposed scheme compared to the optimal scheme is further demonstrated in Fig.4. From which, we can easily identify our scheme indeed approaches the upper bound of the system.

\section{Conclusions}
In this paper, we proposed a joint power allocation and 3D mapping scheme to maximize the system throughput under FD-BS scenario. In detail, we further decompose the joint optimization problem to two sub ones, use the dual method for resolving power allocation problem given each possible pair $(m,n,k)$ then implement 3D mapping given a fixed power allocation. Finally, a low complexity 3D mapping scheme for our model and a semi-closed form solution for power allocation is derived. And the numerical results also show that, our proposed scheme indeed outperforms than the other benchmarks and approaches the performance of optimal scheme.

    \begin{appendices}
	\section{  }

	By applying KKT optimal conditions, the partial derivatives of (22) over $ P_{m,n}^{k,down} $ and $ P_{m,n}^{k,up} $ are set to zeros, as shown by (23) and (24).
	
	\begin{equation}
	\frac{\partial L}{\partial P_{m,n}^{k,down}}=1+{a_{mn}^k}{P_{m,n}^{k,up}}+{P_{m,n}^{k,down}}{a_{bn}^k}-\frac{a_{bn}^k}{\lambda_b}=0
	\end{equation}

	\begin{equation}
	\frac{\partial L}{\partial P_{m,n}^{k,up}}=\frac{a_{mb}^k}{P_{m,n}^{k,up} a_{mb}^k+1}-{\lambda_m}-\frac{P_{m,n}^{k,down} a_{bn}^k a_{mn}^k }{(P_{m,n}^{k,up} a_{mn}^k+1) (P_{m,n}^{k,up} a_{mn}^k+1+P_{m,n}^{k,down} a_{bn}^k)}=0 \\
	\end{equation}
	
	It is generally obtained the optimal variable values by solving a set of equations, i.e. (23) and (24).
	However, this methodology can only be used directly when the feasible regions are differentiable. For our considered problem, it is not capable of utilizing the method directly. This is because when the variable value equals to zero, their derivations are not unique. Thus, we need to discuss them case by case:

	\noindent Case 1: $ P_{m,n}^{k,up} $ and $ P_{m,n}^{k,down} $ are both larger than zero:
	\\

		\begin{equation}
		\begin{split}
		&\left[\lambda_b a_{mb}^k a_{bn}^k-\frac{\lambda_m(a_{bn}^k)^2a_{mb}^k}{a_{mn}^k}\right](P_{m,n}^{k,down})^2+
		\left[\frac{\lambda_m(a_{bn}^k)^2a_{mb}^k}{\lambda_ba_{mn}^k}+(\lambda_ma_{bn}^k-\lambda_ba_{mn}^k)(1+\frac{a_{mb}^ka_{bn}^k}{a_{mn}^k\lambda_b}-\frac{a_{mb}^k}{a_{mn}^k})-a_{mb}^ka_{bn}^k\right]P_{m,n}^{k,down}+\\
		&\left[\frac{a_{mb}^ka_{bn}^k}{\lambda_b}-\frac{\lambda_ma_{bn}^k}{\lambda_b}(1+\frac{a_{mb}^ka_{bn}^k}{a_{mn}^k\lambda_b}-\frac{a_{mb}^k}{a_{mn}^k})\right]=0
		\end{split}
		\end{equation}

	\begin{equation}
	P_{m,n}^{k,up}=\frac{1}{a_{mn}^k}\left(\frac{a_{bn}^k}{\lambda_b}-a_{bn}^kP_{m,n}^{k,down}-1\right)
	\end{equation}
	
	\noindent If (25) and (26) both have positive roots, the problem is feasible, otherwise, infeasible.\\
	
	\noindent Case 2: $ P_{m,n}^{k,down}=0 $, $ P_{m,n}^{k,up}>0 $
	\begin{equation}
	P_{m,n}^{k,up}=\frac{1}{\lambda_m}-\frac{1}{a_{mb}^k}
	\end{equation}
	If $ P_{m,n}^{k,up}>0 $, the problem is feasible, otherwise, infeasible.\\
	
	\noindent Case 3: $ P_{m,n}^{k,down}>0 $, $ P_{m,n}^{k,up}=0 $
	\begin{equation}
	P_{m,n}^{k,down}=\frac{1}{\lambda_b}-\frac{1}{a_{bn}^k}
	\end{equation}
	If $ P_{m,n}^{k,down}>0 $, the problem is feasible, otherwise, infeasible.\\
	
	\noindent Case 4: $ P_{m,n}^{k,down}=0 $, $ P_{m,n}^{k,up}=0 \\\\ $
\end{appendices}

\balance


\begin{thebibliography}{1}
\bibitem{1}
Sabharwal A, Schniter P, Guo D, et al. In-band full-duplex wireless: Challenges and opportunities[J]. IEEE Journal on Selected Areas in Communications, 2014, 32(9): 1637-1652.


\bibitem{2}
Jiang Y, Lau F, Ho I, et al. Resource Allocation for Multi-User OFDMA Hybrid Full-/Half-Duplex Relaying Systems With Direct Links[J]. 2015.

\bibitem{3}
Ng D W K, Schober R. Resource allocation and scheduling in multi-cell OFDMA systems with decode-and-forward relaying[J]. IEEE Transactions on Wireless Communications, 2011, 10(7): 2246-2258.
\bibitem{4}
Zhang J, Li Q, Kim K, et al. On the Performance of Full-duplex Two-way Relay Channels with Spatial Modulation[J]. IEEE Transactions on Communications, 2016: 1-1.

\bibitem{5}
Chae S H, Lim S H. Degrees of freedom of cellular networks: Gain from full-duplex operation at a base station[C]//2014 IEEE Global Communications Conference. IEEE, 2014: 4048-4053.
\bibitem{6}
Zhu F, Gao F, Yao M, et al. Joint information-and jamming-beamforming for physical layer security with full duplex base station[J]. IEEE Transactions on Signal Processing, 2014, 62(24): 6391-6401.
\bibitem{7}
Yu G, Wen D, Qu F. Joint user scheduling and channel allocation for cellular networks with full duplex base stations[J]. IET Communications, 2016, 10(5): 479-486.
\bibitem{8}

Nam C, Joo C, Bahk S. Joint subcarrier assignment and power allocation in full-duplex OFDMA networks[J]. IEEE Transactions on Wireless Communications, 2015, 14(6): 3108-3119.

\bibitem{D2D}
Kim T, Dong M. An iterative hungarian method to joint relay selection and resource allocation for d2d communications[J]. IEEE Wireless Communications Letters, 2014, 3(6): 625-628.

\bibitem{ex}
Lu Z, Shi Y, Wu W, et al. Efficient data retrieval scheduling for multi-channel wireless data broadcast[C]//INFOCOM, 2012 Proceedings IEEE. IEEE, 2012: 891-899.

\bibitem{dual}
Yu W, Lui R. Dual methods for nonconvex spectrum optimization of multicarrier systems[J]. IEEE Transactions on Communications, 2006, 54(7): 1310-1322.

\bibitem{step}
Shor N Z. Minimization methods for non-differentiable functions[M]. Springer, 2012.

\end{thebibliography}
\end{document}